

%
\expandafter\ifx\csname phyzzx\endcsname\relax\else
 \errhelp{Hit <CR> and go ahead.}
 \errmessage{PHYZZX macros are already loaded or input. }
 \endinput \fi
\catcode`\@=11 
\def\spacecheck#1{\dimen@=\pagegoal\advance\dimen@ by -\pagetotal
   \ifdim\dimen@<#1 \ifdim\dimen@>0pt \vfil\break \fi\fi}
\newskip\chapterskip         \chapterskip=\bigskipamount
\newskip\headskip            \headskip=8pt plus 3pt minus 3pt
\newdimen\referenceminspace  \referenceminspace=25pc
\font\fourteenrm=cmr10 scaled\magstep2
\def\refitem#1{\par \hangafter=0 \hangindent=\refindent \Textindent{#1}}
\def\Textindent#1{\noindent\llap{#1\enspace}\ignorespaces}
\def\subspaces@t#1:#2;{
      \baselineskip = \normalbaselineskip
      \multiply\baselineskip by #1 \divide\baselineskip by #2
      \lineskip = \normallineskip
      \multiply\lineskip by #1 \divide\lineskip by #2
      \lineskiplimit = \normallineskiplimit
      \multiply\lineskiplimit by #1 \divide\lineskiplimit by #2
      \parskip = \normalparskip
      \multiply\parskip by #1 \divide\parskip by #2
      \abovedisplayskip = \normaldisplayskip
      \multiply\abovedisplayskip by #1 \divide\abovedisplayskip by #2
      \belowdisplayskip = \abovedisplayskip
      \abovedisplayshortskip = \normaldispshortskip
      \multiply\abovedisplayshortskip by #1
        \divide\abovedisplayshortskip by #2
      \belowdisplayshortskip = \abovedisplayshortskip
      \advance\belowdisplayshortskip by \belowdisplayskip
      \divide\belowdisplayshortskip by 2
      \smallskipamount = \skipregister
      \multiply\smallskipamount by #1 \divide\smallskipamount by #2
      \medskipamount = \smallskipamount \multiply\medskipamount by 2
      \bigskipamount = \smallskipamount \multiply\bigskipamount by 4 }
%
%
%
\newcount\referencecount     \referencecount=0
\newcount\lastrefsbegincount \lastrefsbegincount=0
\newif\ifreferenceopen       \newwrite\referencewrite
\newif\ifrw@trailer
\newdimen\refindent     \refindent=30pt
\def\NPrefmark#1{\attach{\scriptscriptstyle [ #1 ] }}
\let\PRrefmark=\attach
\def\refmark#1{\relax\ifPhysRev\PRrefmark{#1}\else\NPrefmark{#1}\fi}
\def\refend@{\refmark{\number\referencecount}}
\def\refend{\refend@{}\space }
\def\refsend{\refmark{\count255=\referencecount
   \advance\count255 by-\lastrefsbegincount
   \ifcase\count255 \number\referencecount
   \or \number\lastrefsbegincount,\number\referencecount

   \else \number\lastrefsbegincount-\number\referencecount \fi}\space }
\def\refitem#1{\par \hangafter=0 \hangindent=\refindent \Textindent{#1}}
\def\Ref{\rw@trailertrue\REF}
\def\ref{\Ref\?}

\def\REF#1{\r@fstart{#1}%
   \rw@begin{\the\referencecount.}\rw@end}
\def\REFS#1{\r@fstart{#1}%
   \lastrefsbegincount=\referencecount
   \rw@begin{\the\referencecount.}\rw@end}
\def\r@fstart#1{\chardef\rw@write=\referencewrite \let\rw@ending=\refend@
   \ifreferenceopen \else \global\referenceopentrue
   \immediate\openout\referencewrite=referenc.txa
   \toks0={\catcode`\^^M=10}\immediate\write\rw@write{\the\toks0} \fi
   \global\advance\referencecount by 1 \xdef#1{\the\referencecount}}
{\catcode`\^^M=\active %
 \gdef\rw@begin#1{\immediate\write\rw@write{\noexpand\refitem{#1}}%
   \begingroup \catcode`\^^M=\active \let^^M=\relax}%
 \gdef\rw@end#1{\rw@@end #1^^M\rw@terminate \endgroup%
   \ifrw@trailer\rw@ending\global\rw@trailerfalse\fi }%
 \gdef\rw@@end#1^^M{\toks0={#1}\immediate\write\rw@write{\the\toks0}%
   \futurelet\n@xt\rw@test}%
 \gdef\rw@test{\ifx\n@xt\rw@terminate \let\n@xt=\relax%
       \else \let\n@xt=\rw@@end \fi \n@xt}%
}
\let\rw@ending=\relax
\let\rw@terminate=\relax

\def\Closeout#1{\toks0={\catcode`\^^M=5}\immediate\write#1{\the\toks0}%
   \immediate\closeout#1}
%
%
%
%
\newbox\hdbox%
\newcount\hdrows%
\newcount\multispancount%
\newcount\ncase%
\newcount\ncols
\newcount\nrows%
\newcount\nspan%
\newcount\ntemp%
\newdimen\hdsize%
\newdimen\newhdsize%
\newdimen\parasize%
\newdimen\spreadwidth%
\newdimen\thicksize%
\newdimen\thinsize%
\newdimen\tablewidth%
\newif\ifcentertables%
\newif\ifendsize%
\newif\iffirstrow%
\newif\iftableinfo%
\newtoks\dbt%
\newtoks\hdtks%
\newtoks\savetks%
\newtoks\tableLETtokens%
\newtoks\tabletokens%
\newtoks\widthspec%
%
%
\immediate\write15{%
CP SMSG GJMSINK TEXTABLE --> TABLE MACROS V. 851121 JOB = \jobname%
}%
%
%
\tableinfotrue%
\catcode`\@=11
%
%
\def\tstrut{\vrule height3.1ex depth1.2ex width0pt}%
\def\and{\char`\&}
\def\tablerule{\noalign{\hrule height\thinsize depth0pt}}%
\thicksize=1.5pt
\thinsize=0.6pt
\def\thickrule{\noalign{\hrule height\thicksize depth0pt}}%
\def\ctr#1{\hfil\ #1\hfil}%
%
%
%
%
\tablewidth=-\maxdimen%
\spreadwidth=-\maxdimen%
\def\tabskipglue{0pt plus 1fil minus 1fil}%
%
%
\centertablestrue%
%
%
%
%
\parasize=4in%
\gdef\ARGS{########}
\gdef\headerARGS{####}
\def\@mpersand{&}
{\catcode`\|=13
\gdef\letbarzero{\let|0}
\gdef\letbartab{\def|{&&}}%
\gdef\letvbbar{\let\vb|}%
}
{\catcode`\&=4
\def\ampskip{&\omit\hfil&}
\catcode`\&=13
\let&0
\xdef\letampskip{\def&{\ampskip}}%
\gdef\letnovbamp{\let\novb&\let\tab&}
}
\def\begintable{
   \begingroup%
   \catcode`\|=13\letbartab\letvbbar%
   \catcode`\&=13\letampskip\letnovbamp%
   \def\multispan##1{
      \omit \mscount##1%
      \multiply\mscount\tw@\advance\mscount\m@ne%
      \loop\ifnum\mscount>\@ne \sp@n\repeat%
   }
   \def\|{%
      &\omit\widevline&%
   }%
   \ruledtable
}
\long\def\ruledtable#1\endtable{%
%
%
%
   \offinterlineskip
   \tabskip 0pt
   \def\widevline{\vrule width\thicksize}
   \def\endrow{\@mpersand\omit\hfil\crnorm\@mpersand}%
   \def\crthick{\@mpersand\crnorm\thickrule\@mpersand}%
   \def\crthickneg##1{\@mpersand\crnorm\thickrule
          \noalign{{\skip0=##1\vskip-\skip0}}\@mpersand}%
   \def\crnorule{\@mpersand\crnorm\@mpersand}%
   \def\crnoruleneg##1{\@mpersand\crnorm
          \noalign{{\skip0=##1\vskip-\skip0}}\@mpersand}%
   \let\nr=\crnorule
   \def\endtable{\@mpersand\crnorm\thickrule}%
   \let\crnorm=\cr
%
%
   \edef\cr{\@mpersand\crnorm\tablerule\@mpersand}%
   \def\crneg##1{\@mpersand\crnorm\tablerule
          \noalign{{\skip0=##1\vskip-\skip0}}\@mpersand}%
   \let\ctneg=\crthickneg
   \let\nrneg=\crnoruleneg
   \the\tableLETtokens
%
%
   \tabletokens={&#1}
%
%
   \countROWS\tabletokens\into\nrows%
   \countCOLS\tabletokens\into\ncols%
%
%
   \advance\ncols by -1%
   \divide\ncols by 2%
   \advance\nrows by 1%
%
%
   \iftableinfo %
      \immediate\write16{[Nrows=\the\nrows, Ncols=\the\ncols]}%
   \fi%
%
%
   \ifcentertables
      \ifhmode \par\fi
      \line{
      \hss
   \else %
      \hbox{%
   \fi
      \vbox{%
         \makePREAMBLE{\the\ncols}
         \edef\next{\preamble}
         \let\preamble=\next
         \makeTABLE{\preamble}{\tabletokens}
      }
      \ifcentertables \hss}\else }\fi
   \endgroup
   \tablewidth=-\maxdimen
   \spreadwidth=-\maxdimen
}
\def\makeTABLE#1#2{
   {
   \let\ifmath0
   \let\headertab0
   \let\multispan0
%
%
   \ncase=0%
   \ifdim\tablewidth>-\maxdimen \ncase=1\fi%
   \ifdim\spreadwidth>-\maxdimen \ncase=2\fi%
   \relax
%
   \ifcase\ncase %
      \widthspec={}%
   \or %
      \widthspec=\expandafter{\expandafter t\expandafter o%
                 \the\tablewidth}%
   \else %
      \widthspec=\expandafter{\expandafter s\expandafter p\expandafter r%
                 \expandafter e\expandafter a\expandafter d%
                 \the\spreadwidth}%
   \fi %
   \xdef\next{
      \halign\the\widthspec{%
      #1
      \noalign{\hrule height\thicksize depth0pt}
      \the#2\endtable
%
      }
   }
   }
   \next
}
\def\makePREAMBLE#1{
   \ncols=#1
   \begingroup
   \let\ARGS=0
   \edef\xtp{\widevline\ARGS\tabskip\tabskipglue%
   &\ctr{\ARGS}\tstrut}
   \advance\ncols by -1
   \loop
      \ifnum\ncols>0 %
      \advance\ncols by -1%
      \edef\xtp{\xtp&\vrule width\thinsize\ARGS&\ctr{\ARGS}}%
   \repeat
   \xdef\preamble{\xtp&\widevline\ARGS\tabskip0pt%
   \crnorm}
   \endgroup
}
\def\countROWS#1\into#2{
   \let\countREGISTER=#2%
   \countREGISTER=0%
   \expandafter\ROWcount\the#1\endcount%
}%
\def\ROWcount{%
   \afterassignment\subROWcount\let\next= %
}%
\def\subROWcount{%
   \ifx\next\endcount %
      \let\next=\relax%
   \else%
      \ncase=0%
      \ifx\next\cr %
         \global\advance\countREGISTER by 1%
         \ncase=0%
      \fi%
      \ifx\next\endrow %
         \global\advance\countREGISTER by 1%
         \ncase=0%
      \fi%
      \ifx\next\crthick %
         \global\advance\countREGISTER by 1%
         \ncase=0%
      \fi%
      \ifx\next\crnorule %
         \global\advance\countREGISTER by 1%
         \ncase=0%
      \fi%
      \ifx\next\crthickneg %
         \global\advance\countREGISTER by 1%
         \ncase=0%
      \fi%
      \ifx\next\crnoruleneg %
         \global\advance\countREGISTER by 1%
         \ncase=0%
      \fi%
      \ifx\next\crneg %
         \global\advance\countREGISTER by 1%
         \ncase=0%
      \fi%
      \ifx\next\headertab %
         \ncase=1%
      \fi%
      \relax%
      \ifcase\ncase %
         \let\next\ROWcount%
      \or %
         \let\next\argROWskip%
      \else %
      \fi%
   \fi%
   \next%
}
\def\counthdROWS#1\into#2{%
\dvr{10}%
   \let\countREGISTER=#2%
   \countREGISTER=0%
\dvr{11}%
\dvr{13}%
   \expandafter\hdROWcount\the#1\endcount%
\dvr{12}%
}%
\def\hdROWcount{%
   \afterassignment\subhdROWcount\let\next= %
}%
\def\subhdROWcount{%
   \ifx\next\endcount %
      \let\next=\relax%
   \else%
      \ncase=0%
      \ifx\next\cr %
         \global\advance\countREGISTER by 1%
         \ncase=0%
      \fi%
      \ifx\next\endrow %
         \global\advance\countREGISTER by 1%
         \ncase=0%
      \fi%
      \ifx\next\crthick %
         \global\advance\countREGISTER by 1%
         \ncase=0%
      \fi%
      \ifx\next\crnorule %
         \global\advance\countREGISTER by 1%
         \ncase=0%
      \fi%
      \ifx\next\headertab %
         \ncase=1%
      \fi%
\relax%
      \ifcase\ncase %
         \let\next\hdROWcount%
      \or%
         \let\next\arghdROWskip%
      \else %
      \fi%
   \fi%
   \next%
}%
{\catcode`\|=13\letbartab
\gdef\countCOLS#1\into#2{%
   \let\countREGISTER=#2%
   \global\countREGISTER=0%
   \global\multispancount=0%
   \global\firstrowtrue
   \expandafter\COLcount\the#1\endcount%
   \global\advance\countREGISTER by 3%
   \global\advance\countREGISTER by -\multispancount
}%
\gdef\COLcount{%
   \afterassignment\subCOLcount\let\next= %
}%
{\catcode`\&=13%
\gdef\subCOLcount{%
   \ifx\next\endcount %
      \let\next=\relax%
   \else%
      \ncase=0%
      \iffirstrow
         \ifx\next& %
            \global\advance\countREGISTER by 2%
            \ncase=0%
         \fi%
         \ifx\next\span %
            \global\advance\countREGISTER by 1%
            \ncase=0%
         \fi%
         \ifx\next| %
            \global\advance\countREGISTER by 2%
            \ncase=0%
         \fi
         \ifx\next\|
            \global\advance\countREGISTER by 2%
            \ncase=0%
         \fi
         \ifx\next\multispan
            \ncase=1%
            \global\advance\multispancount by 1%
         \fi
         \ifx\next\headertab
            \ncase=2%
         \fi
         \ifx\next\cr       \global\firstrowfalse \fi
         \ifx\next\endrow   \global\firstrowfalse \fi
         \ifx\next\crthick  \global\firstrowfalse \fi
         \ifx\next\crnorule \global\firstrowfalse \fi
         \ifx\next\crnoruleneg \global\firstrowfalse \fi
         \ifx\next\crthickneg  \global\firstrowfalse \fi
         \ifx\next\crneg       \global\firstrowfalse \fi
      \fi
\relax
      \ifcase\ncase %
         \let\next\COLcount%
      \or %
         \let\next\spancount%
      \or %
         \let\next\argCOLskip%
      \else %
      \fi %
   \fi%
   \next%
}%
\gdef\argROWskip#1{%
   \let\next\ROWcount \next%
}
\gdef\arghdROWskip#1{%
   \let\next\ROWcount \next%
}
\gdef\argCOLskip#1{%
   \let\next\COLcount \next%
}
}
}
\def\spancount#1{
   \nspan=#1\multiply\nspan by 2\advance\nspan by -1%
   \global\advance \countREGISTER by \nspan
   \let\next\COLcount \next}%
\def\dvr#1{\relax}%
\def\headertab#1{%
\dvr{1}{\let\cr=\@mpersand%
\hdtks={#1}%
\counthdROWS\hdtks\into\hdrows%
\advance\hdrows by 1%
\ifnum\hdrows=0 \hdrows=1 \fi%
\dvr{5}\makehdPREAMBLE{\the\hdrows}%
\dvr{6}\getHDdimen{#1}%
{\parindent=0pt\hsize=\hdsize{\let\ifmath0%
\xdef\next{\valign{\headerpreamble #1\crnorm}}}\dvr{7}\next\dvr{8}%
}%
}\dvr{2}}
\def\makehdPREAMBLE#1{
\dvr{3}%
\hdrows=#1
{
\let\headerARGS=0%
\let\cr=\crnorm%
\edef\xtp{\vfil\hfil\hbox{\headerARGS}\hfil\vfil}%
\advance\hdrows by -1
\loop
\ifnum\hdrows>0%
\advance\hdrows by -1%
\edef\xtp{\xtp&\vfil\hfil\hbox{\headerARGS}\hfil\vfil}%
\repeat%
\xdef\headerpreamble{\xtp\crcr}%
}
\dvr{4}}
\def\getHDdimen#1{%
\hdsize=0pt%
\getsize#1\cr\end\cr%
}
\def\getsize#1\cr{%
\endsizefalse\savetks={#1}%
\expandafter\lookend\the\savetks\cr%
\relax \ifendsize \let\next\relax \else%
\setbox\hdbox=\hbox{#1}\newhdsize=1.0\wd\hdbox%
\ifdim\newhdsize>\hdsize \hdsize=\newhdsize \fi%
\let\next\getsize \fi%
\next%
}%
\def\lookend{\afterassignment\sublookend\let\looknext= }%
\def\sublookend{\relax%
\ifx\looknext\cr %
\let\looknext\relax \else %
   \relax
   \ifx\looknext\end \global\endsizetrue \fi%
   \let\looknext=\lookend%
    \fi \looknext%
}%
%
%
\def\tablelet#1{%
   \tableLETtokens=\expandafter{\the\tableLETtokens #1}%
}%
\catcode`\@=12

\magnification=\magstep1
\vsize=8.5 true in
\hsize=6.5 true in
\tolerance 10000

\def\markout#1{}
\baselineskip 18pt plus 1pt minus 1pt
\pageno=0
\rightline{\hfil FERMILAB-Pub-93/189-A}
\vskip 0.5in
\centerline{\bf Green's functions for gravitational waves in FRW spacetimes}
\vskip 0.5in
\centerline{R. R. Caldwell}
\vskip 0.5in
\centerline{NASA/Fermilab Astrophysics Center}
\centerline{Fermi National Accelerator Laboratory}
\centerline{P.O. Box 500}
\centerline{Batavia, Illinois 60510-0500}
\vskip 0.2in
$$\eqalign{\rm email: caldwell@virgo.fnal.gov}$$
\vfil
\centerline{\it July, 1993}
\vfil

\baselineskip 12pt plus 1pt minus 1pt

\centerline{\bf Abstract}
\vskip 0.2in
{\noindent\narrower
A method for calculating the retarded Green's function for the
gravitational wave equation in Friedmann-Roberson-Walker spacetimes,
within the formalism of linearized Einstein gravity is developed.
Hadamard's general solution to Cauchy's problem for second-order,
linear partial differential equations is applied to the
FRW gravitational wave equation. The retarded Green's
function may be calculated for any FRW spacetime, with curved or flat
spatial sections, for which the functional form of the
Ricci scalar curvature $R$ is known.
The retarded Green's function for gravitational waves propagating
through a cosmological fluid composed of both radiation and dust is calculated
analytically for the first time. It is also shown that for all FRW spacetimes
in which the Ricci scalar curvatures does not vanish, $R \neq 0$, the Green's
function violates Huygens' principle; the Green's function has support
inside the light-cone due to the scatter of gravitational
waves off the background curvature.}
\eject
\baselineskip 12pt plus 1pt minus 1pt

\REF\STEBBINSA{
S. Veeraraghavan and A. Stebbins, Astrophys. J. {\bf 365}, 37 (1990).}
\REF\WAYLENA{
P. C. Waylen, Proc. R. Soc. London A. {\bf 362}, 233 (1978).}
\REF\WAYLENB{
P. C. Waylen, Proc. R. Soc. London A. {\bf 362}, 245 (1978).}
\REF\HADAMARD{
J. Hadamard, {\it Lectures on Cauchy's problem in linear partial differential
equations},
Dover Publications: New York (1952).}
\REF\DEWITTA{
Bryce S. DeWitt and Robert W. Brehme, Annals of Physics {\bf 9}, 220 (1960).}
\REF\DEWITTB{
Bryce S. DeWitt, {\it Dynamical theory of groups and fields}, Gordon and
Breach: New York (1965).}
\REF\MCLENAGHAN{
R. G. McLenaghan, Proc. Camb. Phil. Soc. {\bf 65}, 139 (1969).}
\REF\WEINBERG{
S. Weinberg, {\it Gravitation and Cosmology}, Wiley: New York (1972).}
\REF\MTW{
C. Misner, K. Thorne and J. Wheeler, {\it Gravitation}, W. H. Freeman: San
Francisco (1973).}
\REF\GRISHCHUKB{
L. P. Grishchuk and A. D. Popova, Sov. Phys.-JETP {\bf 53}, 1 (1981).}
\REF\GRISHCHUKC{
L. P. Grishchuk and A. D. Popova, J. Phys. A: Math. Gen. {\bf 15}, 3525
(1982).}
\REF\BALLENA{
Bruce Allen and Theodore Jacobson, Commun. Math. Phys. {\bf 103}, 669 (1986).}
\REF\BALLENB{
B. Allen, Nuc. Phys. {\bf B287}, 743 (1987).}
\REF\TUROK{
N. Turok, Phys. Rev. Lett. {\bf 63}, 2652 (1989).}
\REF\PARKERA{
L. Parker, Phys. Rev. {\bf 183}, 1057 (1969).}
\REF\BIRRELL{
Birrell and Davies, {\it Quantum fields in curved space}, Cambridge University
Press:
Cambridge (1986).}
\REF\DOWKERA{
J. S. Dowker, Annals of Physics {\bf 62}, 361 (1971).}
\REF\DOWKERB{
J. S. Dowker, Annals of Physics {\bf 71}, 577 (1972).}
\REF\GRISHCHUKA{
L. P. Grishchuk, Sov. Phys.-JETP {\bf 40}, 409 (1974).}
\REF\GRISHCHUKD{
L. P. Grishchuk, Sov. Phys. Usp. {\bf 31}, 940 (1988).}
\REF\STEBBINSB{
Albert Stebbins and Shoba Veeraraghavan, ``MBR anisotropy from scalar field
gradients'',
Fermilab-pub-92/188-A (1992).}
\REF\BENNETT{
D. Bennett and S. H. Rhie, UCRL-JC-111244 (1992).}
\REF\TURNERA{
Michael S. Turner, ``On the production of scalar and tensor perturbations
in inflationary models'', Fermilab-pub-93/026-A (1993).}
\REF\TURNERB{
A. Kosowsky, Michael S. Turner, and R. Watkins, Phys. Rev. Lett. {\bf 69}, 2026
(1992).}
\REF\MATHEMATICA{
Wolfram Research, Inc., {\rm MATHEMATICA} (Wolfram Research, Inc.,
Champaign, Illinois, 1992).}
\REF\MATHTENSOR{
L. Parker and S. M Christensen, {\rm MATHTENSOR} (MathSolutions, Inc., Chapel
Hill, North Carolina, 1992).}

\centerline{\bf I. Introduction}

The Green's function is an important tool for evaluating the solution
to a partial differential equation for a given set of boundary conditions.
The Green's function represents the solution to Cauchy's problem:
given a set of properly formulated initial data, the Green's function
may be used to uniquely determine the final, evolved data. In physics, one may
determine
the evolution of a field according to its equations of motion, given some
initial data. The techniques used in this paper have been widely applied
to the calculation of Green's functions, or propagators, for quantum fields
in curved spacetime. Here, those techniques will be applied to classical
fields in cosmology. In this paper I will determine the evolution of a tensor
field,
representing gravitational waves, given a source stress-energy tensor.
The goal of this paper, then, is to calculate the Green's function for
gravitational waves evolving in an arbitrary FRW spacetime.

The Green's functions may be useful for the study of gravitational
radiation in FRW spacetimes, especially as it relates to observational
cosmology (for an example, see [\STEBBINSA]). For situations in which the
stress-energy tensor of a source of gravitational waves is known, the Green's
function may be used to calculate the properties of the tensor metric
perturbations, and the spectrum of gravitational radiation.  Examples of
such sources are cosmic strings or scalar fields present in the early universe.
The tensor perturbations generated by such sources may be manifest as
anisotropies
in the cosmic microwave background, or as a spectrum of stochastic
gravitational radiation. Then, the Green's function may be used for the
calculation of physically observable quantities.

The organization of the paper is as follows. In section II, the formalism
for calculating the Green's function, following Hadamard's general
solution to Cauchy's problem for the wave equation, will be
presented. A similar technique was developed by Waylen in [\WAYLENA,\WAYLENB].
In section III, this technique will be applied to find the
Green's functions for arbitrary FRW spacetimes. The main result of this paper,
a general expression for the Green's function for any FRW spacetime
in which $R$ is known, will be presented.
Several specific cases will
be evaluated explicitly, including the case in
which the cosmological fluid undergoes
a smooth transition from radiation- to dust-dominated expansion. In section IV
the properties and applications of these Green's functions will be discussed.
The focus will be on the cosmological significance of those terms which violate
Huygens' principle. A final summary of this work will be presented in section
V.

\vskip 0.2in
\centerline{\bf II. Formalism for calculating Green's functions}

The technique for calculating Green's functions follows from Hadamard's
general solution to Cauchy's problem for a second-order, linear partial
differential equation [\HADAMARD]. The most general
form for the solution to the tensor wave equation
$${h_{ab;c}}^{c} + \Lambda h_{ab} = 0 \eqno(II.1)$$
may be written as an expansion in powers of the non-local, biscalar of geodetic
interval $\sigma(x,x_i) = {1 \over 2} s(x,x_i)^2$, where $s(x,x_i)$ is the
geodetic interval
between a fixed 4-vector $x_i$ and a free 4-vector $x$ in the spacetime.
(Note that $\Lambda$ is some arbitrary function, {\it not} the cosmological
constant.)
The Green's function solution is a bitensor, an object with tensor indices on
the
spatial hypersurfaces $x_i$ and $x$:
$${G(x,x_i)_{ab}}^{{c_i}{d_i}} = {1 \over 4 \pi^2}\bigl[{u_{ab}}^{{c_i}{d_i}}
{1 \over \sigma} +
{v_{ab}}^{{c_i}{d_i}} \log{|\sigma|} + {w_{ab}}^{{c_i}{d_i}}\bigr].
\eqno(II.2)$$
Here, $u$, $v$, and $w$ are bitensors which are free of singularities.
Extending $II.2$ into the complex-$\sigma$ plane, and retaining only the
imaginary part, we may obtain the retarded solution [\DEWITTA,\DEWITTB]
$${G^{ret}(x,x_i)_{ab}}^{{c_i}{d_i}} = {1 \over 4
\pi}\bigl[{u_{ab}}^{{c_i}{d_i}} \delta(\sigma) -
{v_{ab}}^{{c_i}{d_i}} \theta(-\sigma)\bigr] \theta(\lambda). \eqno(II.3)$$
The parameter $\lambda$ runs between the spacetime points $x_i$ and $x$,
such as the time, and is positive when the spacelike hypersurface containing
$x$ lies to the future of that containing $x_i$. In that this equation
represents
the solution to the wave equation which is homogeneous everywhere except at a
single point, we may obtain the solution to the inhomogeneous wave equation
$${h _{ab;c}}^{c} + \Lambda h_{ab} = f_{ab}\eqno(II.4)$$
by summing the contribution of many points:
$$h(x)_{ab} = \int_{x_i}^x \sqrt{-g(x')}
d^4 x' G^{ret}{(x,x')_{ab}}^{{c_i}{d_i}} f(x')_{{c_i}{d_i}}. \eqno(II.5)$$
Thus, the Green's function may be used to construct solutions to the wave
equation.
Equation $II.3$, then, gives the retarded Green's function which we seek.

A prescription for calculating the non-singular, bitensor quantities
$u$ and $v$ may be obtained by generalizing the work of DeWitt and Brehme
[\DEWITTA,\DEWITTB]
and McLenaghan [\MCLENAGHAN] on the scalar and vector wave equations.
Such a calculation has been carried out by Waylen [\WAYLENA,\WAYLENB].
One may apply Hadamard's general solution $II.2$
to the homogenous wave equation, and equate terms with matching powers
of $\sigma$. One obtains
$$\eqalign{ ({\sigma_{;c}}^{c} - 4) {u_{ab}}^{{c_i}{d_i}} +
2 {{u_{ab}}^{{c_i}{d_i}}}_{;c} \sigma^{;c} &= 0 \cr
2 \sigma_{;c} {v_{ab}}^{{c_i}{d_i};c} +
({\sigma_{;c}}^{c} - 2){v_{ab}}^{{c_i}{d_i}} +
{u_{ab}}^{{c_i}{d_i}} {{}_{;c}}^c + \Lambda {u_{ab}}^{{c_i}{d_i}} &= 0 . \cr}
\eqno(II.6)$$
Manipulating these equations, we find that [\MCLENAGHAN]
$$\eqalign{ \ln{u} =& -{1 \over 2} \int {d\lambda \over
\lambda}({\sigma_{;c}}^{c} - 4) \cr
{v_{ab}}^{{c_i}{d_i}} =& -{1 \over 2 \lambda} {u_{ab}}^{{c_i}{d_i}}
\int d\lambda ({u_{ef}}^{{g_i}{h_i}}{{}_{;c}}^{c}
+ \Lambda {u_{ef}}^{{g_i}{h_i}}) {(u^{-1})^{ef}}_{{g_i}{h_i}}. \cr}
\eqno(II.7)$$
In general, the bitensors $u$ and $v$ will be proportional
to the bitensor of parallel geodetic transport (see the discussion
of such bitensors in [\DEWITTA]).
The functions $u$ and $v$ may be evaluated once the
biscalar of geodetic interval $\sigma$ is known.
Therefore, for a given metric, $\sigma$, $u$, and $v$ may be determined, and
the
Green's function for the tensor wave equation may be evaluated. This general
technique may be applied to the wave equation in any spacetime.

\vskip 0.2in
\centerline{\bf III. Green's functions in FRW spacetimes}

The gravitational wave equation for which the Green's function
will be determined, describing the propagation of
tensor metric perturbations on a background spacetime, may be derived
in the context of linearized, Einstein gravity. (For some details on
linearized,
Einstein gravity, see [\WEINBERG,\MTW]).
I will consider tensor metric perturbations, $g_{ab} \rightarrow g_{ab} +
h_{ab}$,
where the line element for the background, FRW metric is
$$ds^2 = a(\eta)^2(- d\eta^2 + {dr^2 \over 1 - K r^2} + r^2 d\theta^2
+ r^2 \sin^2\theta d\phi^2). \eqno(III.1)$$
Due to the symmetries of the FRW spacetime, I may apply the transverse,
traceless, synchronous (TTS) gauge conditions [\GRISHCHUKB,\GRISHCHUKC]
$${h_{ab}}^{;b} = h_{ab} g^{ab} = h_{ab} t^b = 0 \eqno(III.2)$$
to isolate the physical degrees of freedom in the gravitational radiation.
Here, $t_a$ is a unit 4-vector pointing in the time direction $\eta$,
orthogonal
to spacelike hypersurfaces.
Ultimately, the gravitational wave equation is found to be
$$ 16 \pi G  T^{source}_{ab} = {h_{ab;c}}^c
+ 2 {{{R^c}_{a}}^d}_b h_{cd}. \eqno(III.3)$$
Here, ${{{R^c}_{a}}^d}_b$ is
the Riemann curvature tensor in the background FRW spacetime,
and $G$ is the gravitational contant.
(We use units such that the speed of light is unity: $c=1$.)
The stress-energy tensor $T^{source}_{ab}$ represents the external source of
the gravitational waves.

We may now apply the techniques developed in the previous section
in order to evaluate the retarded Green's function.
The bitensor functions $u$ and $v$ are proportional
to the transverse, traceless, synchronous bitensor
${\Sigma(x,x_i)_{ab}}^{{c_i}{d_i}}$, which projects only the
TTS portion of the source stress-energy tensor from the spacelike hypersurface
at $x_i$ to $x$.
The TTS bitensor ${\Sigma_{ab}}^{{c_i}{d_i}}$ may be constructed through the
use of
the spatial biscalar of geodetic interval,
$\mu(\vec x, \vec x_i) = |\vec x - \vec x_i|$,
following the work of Allen and Jacobson [\BALLENA,\BALLENB].
Defining
$$\eqalign{V_a = \nabla_a \mu(\vec x, \vec x_i)
&\qquad P_{ab} = g_{ab} + t_a t_b \cr
V_{a_i} = \nabla_{a_i}\mu(\vec x, \vec x_i)
&\qquad P_{a{b_i}} = {g^b}_{b_i} P_{ab} \cr}
\eqno(III.4)$$
the unique TTS bitensor with indices $a,b$ on the tangent space
at $x$, and $c_i,d_i$ at $x_i$ is
$$\eqalign{{\Sigma_{ab}}^{{c_i}{d_i}} =&  {P_a}^{c_i}{P_b}^{d_i} +
{P_a}^{d_i}{P_b}^{c_i} - P_{ab} P^{{c_i}{d_i}}
+ V_aV_b P^{{c_i}{d_i}} + P_{ab} V^{c_i}V^{d_i} \cr
& + {P_a}^{c_i} V_b V^{d_i} + {P_a}^{d_i} V_b V^{c_i}
+ {P_b}^{c_i} V_a V^{d_i} + {P_b}^{d_i} V_a V^{c_i} +
V_a V_b V^{c_i} V^{d_i}.\cr} \eqno(III.5)$$
Here, $V^a {\Sigma_{ab}}^{{c_i}{d_i}} = g^{ab} {\Sigma_{ab}}^{{c_i}{d_i}} =
t^a {\Sigma_{ab}}^{{c_i}{d_i}} = 0$, such that it is
transverse, traceless, and synchronous.
Thus, combining equations $II.3,7$, and $III.3$ the entire prescription
for calculating the Green's function for tensor metric perturbations is
specified.

The calculation of the Green's function for FRW spacetimes may now be
carried out. Having specified the background metric, the biscalar of geodetic
interval is
$\sigma = {1 \over 2}(|\vec x - \vec x_i|^2 - (\eta - \eta_i)^2)$.
Evaluating equations $II.7$,
$$\eqalign{{u_{ab}}^{{c_i}{d_i}} =& {C(|\vec x - \vec x'|/ {\cal R}) \over
a(\eta_i) a(\eta)}{\Sigma_{ab}}^{{c_i}{d_i}}\cr
{v_{ab}}^{{c_i}{d_i}} =& -{{u_{ab}}^{{c_i}{d_i}} \over 12  (\eta - \eta_i)}
\int_{\eta_i}^{\eta} a^2(\eta) R d \eta . \cr} \eqno(III.6)$$
In this equation
$C(x) = x/\sin(x)$ and ${\cal R}$ is the curvature scale
of the spatial sections of the spacetime such that $^3R = 6 {\cal R}^{-2}$.
In the case $K=0$, ${\cal R} \rightarrow \infty$ so that $C=1$.
Applying the above results to equations $II.3$ and $III.3$, one may write the
Green's function as
$$\eqalign{ {G^{ret}(x,x_i)_{ab}}^{{c_i}{d_i}} =& {C(|\vec x - \vec x'|/ {\cal
R}) \over 4 \pi
a(\eta_i) a(\eta)} \Bigl[
{\delta(\eta - \eta_i - |\vec x - \vec x_i|)
\over |\vec x - \vec x_i|}\cr
+ {1 \over 12} & \bigl(\int_{\eta_i}^{\eta} a^2(\eta) R d\eta \bigr)
{\theta(\eta - \eta_i - |\vec x - \vec x_i|) \over \eta - \eta_i} \Bigr]
\theta(\eta-\eta_i) {\Sigma(x,x_i)_{ab}}^{{c_i}{d_i}}. \cr } \eqno(III.7)$$
Note that the Green's function is symmetric under interchange $x
\leftrightarrow x_i$,
as would be expected. Also, the Green's function possesses the
correct dimensionality, $({\rm length})^{-2}$, where the coordinates
$(\eta,\vec x)$
are dimensionless and the expansion scale factor $a$ carries units of length.
This equation composes the main result of this paper.

The Green's function may be evaluated for a number of specific cases.
The results are displayed in the accompanying table.
({\it See table}.)
In this table, $\alpha = -1,1,2$ for deSitter spacetime,
radiation- and dust-dominated
expansion, respectively. Here, the equation of state of the
cosmological fluid is $p = {2 - \alpha \over 3 \alpha} \rho$.
Minkowski spacetime is given by $K=0$ and $a(\eta)=1$
(for which case the preceding equation of state is invalid).
The first three entries in the table give the expansion scale factor
and the Green's function for an ideal cosmological fluid, in a spacetime
with flat and curved spatial sections. The results from the first entry,
for flat spatial sections, have been previously calculated for the specific
cases $\alpha = -1,1,2$ [\STEBBINSA,\TUROK].
The Green's function for
arbitrary $\alpha$ and for any $K$
is given here for the first time.
The last three entries in the table give the expansion scale factor and Green's
function for a mixed cosmological fluid, in a spacetime with flat and curved
spatial
sections. In the case of the mixed ideal fluid, the energy density is
$\rho = \rho_{rad} + \rho_{mat}$ and the pressure is $p = {1 \over 3}
\rho_{rad}$,
where  $\tau$ is a function of $\rho_{crit}(\eta_0)$ and $\rho_{rad}(\eta_0)$,
and $\eta_0$ is the present-day conformal time.
Then, for $\eta \ll \tau$, the fluid is radiation-dominated. For
$\eta \gg \tau$, the fluid is dust-dominated. The Green's function for this
mixed case
is presented here for the first time. This case represents
the most realistic cosmological scenario.

\eject
\centerline{\bf IV. Properties and applications of Green's functions}

The Green's functions describing the evolution of gravitational radiation
in FRW spacetimes, derived in the preceding section, display interesting
properties
and have useful applications, which may now be discussed. Specifically, one may
examine the physical consequences of the terms in the Green's functions
which contribute to the violation of Huygens' principle. Applications of the
analytic
Green's functions for open and closed spacetimes, for a combined radiation and
matter fluid will be discussed.

Huygens' principle states that the effect of a luminous disturbance about some
point at time $t$ will be localized at a later time $t'$
in a very thin spherical shell at a radius $c (t'-t)$ [\HADAMARD].
So, a wave packet travels spherically outward on the light cone from some
source.
{}From examining equation $II.3$, one finds that for $v \neq 0$,
the Green's function has support inside the light cone; the ``wave packet''
disperses
as it travels and is not localized on a thin spherical shell.
In an FRW spacetime, equation $III.7$ indicates that when $R \neq 0$
Huygens' principle is violated.

The violation of Huygens' principle is due to the coupling of the curvature
of spacetime to the tensor metric perturbations. The energy in the
background curvature feeds into the gravitational radiation. This coupling
of spacetime curvature to fields has been well studied on the microscopic level
in the context of particle creation [\PARKERA,\BIRRELL].
(See references [\DOWKERA,\DOWKERB] for a discussion of quantum mechanical
propagators
with respect to Huygens' principle.)
A classical analogue of particle
creation is observed here on the macroscopic level [\GRISHCHUKA,\GRISHCHUKD].
Examining the table, the function $V$ increases with the expansion $a(\eta)$;
$V$ represents
the amplification of the classical gravitational field due
to the expansion. One may say that energy is transferred from the expansion to
the
gravitational field.
Ultimately, it may be simply stated that
Huygens' principle for gravitational waves propagating in an FRW spacetime
is satisfied only when $R=0$, or the spacetime is filled
with a conformally-invariant radiation fluid.

The Green's function for the case of the mixed radiation-plus-dust fluid,
for $K = \pm 1,0$, as presented in the table, is new. These analytic
expressions
should be useful for examining the evolution of gravitational waves through
the transition from radiation- to dust-dominated expansion.

The primary application of the Green's functions derived in this paper is
for the situation in which the source stress-energy tensor is known.
The source ought to be ``stiff'', in that the source evolves freely of the
background
spacetime and the perturbations it produces.
Some examples are cosmic strings, global topological defects, and scalar fields
[\STEBBINSA,\TUROK,\STEBBINSB,\BENNETT]. Of interest are the anisotropies
produced in the microwave background, and the spectrum of gravitational
radiation
emitted.

The anisotropies in the microwave background may be calculated by summing the
contributions to the temperature fluctuations caused by the perturbations in
the gravitational field along the path length of a photon traveling from the
surface of last scattering.
$$\eqalign{
{\delta T \over T} =& -{1 \over 2} \int d\lambda
e^a(\lambda) e^b(\lambda) {\partial \over \partial \eta}
\bigl({1 \over a^2(\eta)} h_{ab} \bigr) \cr
=& -{1\over 2} \int d\lambda
e^a(\lambda) e^b(\lambda) {\partial \over \partial \eta}
\Bigl({1\over a^2(\eta)}
\int \sqrt{-g_i} d^4x_i {G^{ret}(x,x_i)_{ab}}^{{c_i}{d_i}} 16 \pi G
T^{source}(x_i)_{{c_i}{d_i}} \Bigr) \cr} \eqno(IV.1)$$
For a given source stress-energy tensor, the temperature fluctuations
$\delta T/T$ causedd by cosmic strings, global topological
defects, or scalar fields [\STEBBINSA,\STEBBINSB,\TUROK,\BENNETT] may be
calculated.

The energy density and power in the gravitational radiation emitted by a given
source may be examined by calculating the stress-energy tensor
$$ \eqalign{8 \pi G  T^{grav}_{ab} =& {1 \over 2}h^{cd}\bigl[h_{da;bc} +
h_{db;ac}
- h_{ab;cd} - h_{cd;ab} \bigr] - {1 \over 4}{h^c}_{d;a} {h^d}_{c;b} \cr
&+ {1 \over 2} g_{ab} \bigl[ h^{cd} {h_{cd;n}}^n + {1 \over 4} h_{cd;n}
h^{cd;n}
- h^{cd} h_{cd} \bigl({K \over a^2} + {{a'}^2 \over a^4} \bigr) \bigr]. \cr
}\eqno(IV.2)$$
This stress-energy tensor, the terms occuring in the perturbed
Einstein's equations which are
second-order in $h$, may be evaluated for a given source through the use of the
Green's function. This expression may have use in studying the power in tensor
metric perturbations produced during inflation
(for a recent review, see [\TURNERA]), or through the collision
of bubbles formed in a first-order phase transition [\TURNERB].
In this case, the Green's functions
derived in the preceding section for the mixed ideal fluid may be especially
useful for
calculating the evolution of the gravitational waves through the transition
from
radiation- to dust-dominated expansion.

Finally, it is interesting to note that while the Green's function
for the mixed ideal fluid
has been obtained, the mode functions for the gravitational waves in such
a case apparently cannot be written in terms of known functions.
In the case of a single, ideal fluid background, the mode
functions, solutions to equation $II.1$, take the
form of Bessel functions. The case of the mixed fluid, however,
appears much more complicated.

\vskip 0.2in
\centerline{\bf V. Conclusion}

In this paper the retarded Green's functions for the
gravitational wave equation in Friedmann-Robertson-Walker spacetimes,
within the formalism of linearized Einstein gravity, were calculated.
While the form for the Green's function for a generic FRW spacetime was
presented,
several specific cases were considered. These cases, presented
for the first time, include an ideal cosmological
fluid with an equation of state $p = {2 - \alpha \over 3 \alpha} \rho$,
and the case of a mixed ideal fluid of radiation and collisionless dust,
where $\rho = \rho_{rad} + \rho_{mat}$ and $p = {1 \over 3} \rho_{rad}$.
The Green's functions for varying spatial curvature, $K = 0, \pm1$ were
also considered. It was also shown that for all non-conformally invariant
FRW spacetimes, in which $R \neq 0$, the Green's function
violates Huygens' principle.
This is the classical analogue of particle creation in a varying gravitational
field, as the gravitational waves scatter off the background curvature and gain
energy from the cosmological expansion. Finally, it was indicated how these
Green's functions may be applied to the calculation of the microwave
anisotropies
and spectrum of gravitational radiation produced by ``stiff'' sources.
These applications of the Green's functions will be carried out in a future
work.

\vfil
\eject
\centerline{\bf Acknowledgements}

This work was supported in part by the DOE and the NASA (grant $\#$ NAGW-2381)
at Fermilab. I would like to thank Bruce Allen and Leonid Grishchuk for helpful
conversations during the course of this investigation. Some of the work
presented in this paper was carried out using Mathematica [\MATHEMATICA] and
MathTensor [\MATHTENSOR].

\vskip 0.5in
\centerline{\bf Table}
\vskip 0.2in

\def\tstrut{\vrule height 4.0ex depth 1.8ex width 0pt}

\begintable
\multispan{5}\tstrut
\hfil {\bf Green's functions for gravitational waves in FRW spacetimes} \hfil
\crthick
\multispan{5}\tstrut
\hfil
${G^{ret}(x,x_i)_{ab}}^{{c_i}{d_i}} = {C(\vert\vec x - \vec x_i\vert/ {\cal R})
\over 4 \pi a(\eta_i) a(\eta)}
\bigl[{\delta(\eta - \eta_i - \vert\vec x - \vec x_i\vert)
\over \vert\vec x - \vec x_i\vert} + V
{\theta(\eta - \eta_i - \vert\vec x - \vec x_i\vert) \over {\eta -
\eta_i}}\bigr]
\theta(\eta-\eta_i) {\Sigma(x,x_i)_{ab}}^{{c_i}{d_i}}$
\hfil
\crthick
${\bf K} $ |
${\bf C(\vert\vec x - \vec x_i\vert/ {\cal R})} $ |
${\bf a(\eta)}$ |
${\bf V(\eta)}$ |
{\bf cosmology} \crthick
$0$ |
$1$ |
$a(\eta_i)[{\eta \over \eta_i}]^\alpha$ |
${\alpha(\alpha-1) ({\eta - \eta_i}) \over 2 \eta_i \eta}$ |
ideal fluid \cr
$-1$ |
${\vert\vec x - \vec x_i\vert \over {\cal R} }/ \sinh({\vert\vec x - \vec
x_i\vert \over {\cal R}})$ |
$a(\eta_i)[{\sinh(\eta/\alpha) \over \sinh(\eta_i/\alpha)}]^\alpha$ |
${1 \over 2}(\alpha-1)\bigl({\sinh[(\eta-\eta_i)/\alpha] \over
\sinh(\eta/\alpha) \sinh(\eta_i/\alpha)}\bigr)$ |
ideal fluid \cr
$1$ |
${\vert\vec x - \vec x_i\vert \over {\cal R} }/ \sin({\vert\vec x - \vec
x_i\vert \over {\cal R}})$ |
$a(\eta_i)[{\sin(\eta/\alpha) \over \sin(\eta_i/\alpha)}]^\alpha$ |
${1 \over 2}(\alpha-1)\bigl({\sin[(\eta-\eta_i)/\alpha] \over \sin(\eta/\alpha)
\sin(\eta_i/\alpha)}\bigr)$ |
ideal fluid \cr
$0$ |
$1$ |
$a(\eta_i) [{\eta(\eta+\tau) \over \eta_i(\eta_i+\tau)}]$ |
${1 \over \tau} \log{\eta(\eta_i+\tau) \over \eta_i(\eta+\tau)}$ |
mixed fluid \cr
$-1$ |
${\vert\vec x - \vec x_i\vert \over {\cal R} }/ \sinh({\vert\vec x - \vec
x_i\vert \over {\cal R}})$ |
$a(\eta_i)[{\tau \sinh\eta + \sinh^2(\eta/2) \over \tau \sinh\eta_i +
\sinh^2(\eta_i/2)}]$ |
${1 \over 4 \tau} \log\bigl[ {2 \tau \coth(\eta_i/2) + 1 \over 2 \tau
\coth(\eta/2) + 1} \bigr]$ |
mixed fluid \cr
$1$ |
${\vert\vec x - \vec x_i\vert \over {\cal R} }/ \sin({\vert\vec x - \vec
x_i\vert \over {\cal R}})$ |
$a(\eta_i)[{\tau \sin\eta + \sin^2(\eta/2) \over \tau \sin\eta_i +
\sin^2(\eta_i/2)}]$ |
${1 \over 4 \tau} \log\bigl[ {2 \tau \cot(\eta_i/2) + 1 \over 2 \tau
\cot(\eta/2) + 1} \bigr]$ |
mixed fluid
\endtable
\vskip 0.2in

\par\penalty-400\vskip\chapterskip\spacecheck\referenceminspace
   \ifreferenceopen \Closeout\referencewrite \referenceopenfalse \fi
   \line{\fourteenrm\hfil REFERENCES\hfil}\vskip\headskip
   \input referenc.txa
   
\end